**Title: Nanostructured Electrodes as Random Arrays of Active Sites: Modeling and Theoretical Characterization**

Alexander Oleinick[1,z], Oleksii Sliusarenko[1], Irina Svir[1,2], Christian Amatore[1,3*z]

[1] CNRS – Ecole Normale Superieure, PSL Research University – Sorbonne University, UMR 8640 "PASTEUR", 24 rue Lhomond, 75005 Paris, France

[2] Design Automation Department, Kharkiv National University of Radioelectronics, 14 Nauky avenue, 61166 Kharkiv, Ukraine

[3] State Key Laboratory of Physical Chemistry of Solid Surfaces, College of Chemistry and Chemical Engineering, Xiamen University, 361005 Xiamen, China.

[z] oleksandr.oliynyk@ens.fr; christian.amatore@ens.fr

**Abstract Text**

This review presents the main principles underlying the theoretical description of the behavior of regular and random arrays of nanometric active sites. It is further shown how they can be applied for establishing a useful semi-analytical approximation of the arrays responses under diffusion limited conditions when they involve the common situation of active sites with identical sizes. This approximation is general and, as exemplified for different type of arrays, can be employed for describing the behavior of any array involving arbitrary distributions of their active sites onto the substrate surface. Furthermore, this efficient approach allows statistical characterization of active sites distributions of any array based on chronoamperometric data.

**Introduction Text**

Nanostructured electrochemical interfaces (NECI) are frequently encountered in practice. For instance, partially blocked electrodes [1], defects and pinholes in self-assembled monolayers [2], pores in synthetic or natural membranes [3, 4], catalytic nanoparticles (NP) or any type of catalytic site distributed at an inert or less reactive substrate [5] etc. are physical representation of NECI. In these systems, the location of the active sites cannot be controlled (excluding some specific situations) thus leading to disordered surface structures referred as random arrays of active sites. In order to model properly the response of such systems one has to deal simultaneously with multiple scales: electron transfer events at the nano- or microsized active site surface, varying distances between the disordered active sites and interaction of the diffusion layers originated at each sites, global diffusion layer propagation towards the bulk solution, eventual homogeneous kinetics etc. In addition, theoretical description of random arrays is complicated by the fact that taking into account exact positions of the active sites with respect to each other is either difficult or impossible, however, the electrochemical response of the system can be modeled employing statistical information about active sites dispersal.

One of the earliest attempts to model disordered systems in the context of partially blocked surfaces was made by one of us [1]. In this work, the behaviors of random arrays with disk- or band-like electrochemically active surface structures were investigated by relying on regular arrays with uniform size of active sites. Such treatment simplified greatly description of the system, but at the same time, it has a good predictive power as confirmed by notable citation of this early work. Nevertheless, due to computer limitations at the time it was difficult to do otherwise. The situation is different nowadays and more sophisticated approaches can be used as detailed below.

**Experimental**

All the results referred in this work were obtained with home-build programs (Python, Delphi) or using Mathematica 12.0 [6]. The software were run at PC with Intel Xeon E5-2650 v2 processor.

**Results**

We wish to take advantage of this review to summarize and unify a series of our previous works in which we considered an archetypical situation of arrays of nano- or microdisk active sites embedded in the inert surface of a substrate according to an arbitrary probability distribution. Such systems bear the most general characteristics of the systems mentioned in the Introduction. It should be noted, however, that all considerations herein are restricted to the case of active sites with monodisperse size distribution performing under diffusion limited conditions.

Modelling and simulating such disordered systems is a complex problem, however it will be shown that it can be performed on the basis of principles similar to those used for theoretical description of the regular arrays (e.g. arrays with the active sites located on a squared or hexagonal lattice).

*Construction of the unit cells.* One of the main features of the regular arrays is their symmetry. Due to the symmetry the diffusion layers around each active site are identical (except those on the edge of the array, vide infra) both at short times after start of the experiment when diffusion layers of neighboring sites are separate and do not interact, but also at intermediate and long times when the diffusion layers are partially or fully overlap [1, 7]. This allows delineating virtual 3D-spaces around each active site, in which all individual diffusion layers behave in an identical way. These spaces are called unit cell and their shape depends on the array lattice (Fig.1a). The limiting planes of the unit cells are the symmetry planes between corresponding neighboring sites, hence the net flux across these boundaries is zero. This significantly simplifies modeling of the regular arrays since it permits to consider each unit cell as performing independently from the rest of array and model mass transport and kinetics inside a cell without influence of its identical neighbors. Hence the total current of the regular array, $i_{array}$, can be written as:

$$i_{reg.arr.}(t) = N_{site}\, i_{cell}(t) \qquad (1)$$

where $N_{site}$ is the number of active sites in the array and $i_{cell}$ is a current contribution of a single unit cell.

Two points should be noted with respect to the previous paragraph. First, strictly speaking Eq.(1) is valid only for an infinite regular array or practically for a sufficiently large array with number of active sites at the edge of the array being much smaller than the number of the sites at the interior of the array. Otherwise, in order to take properly into account contribution of the edge active sites to overall array current a correction similar to those used in [8] should be applied. Second, zero net flux boundary conditions are assigned at the unit cells boundaries making them acting independently from each other. However, the virtual boundary of the unit cells are just useful theoretical constructions and, evidently, there exist exchange of the molecules between the neighboring unit cells. Though, these exchanges are fully compensated due to the symmetry of the diffusion layers in adjacent unit cells which results in a net zero total flux across the inter-cell boundary.

Similar construct can be done in the case of a disordered array via Voronoi tessellation [9-12]. This technique is used in various branches of natural sciences and represents a natural generalization of the approach used for regular arrays. Indeed, in Voronoi tesselation symmetry planes are constructed between the closest active sites as discussed above, but due to the random location of the neighbour sites this results in a non-uniform unit cells, which are then named Voronoi cells. As can be seen in Fig.1b each segment of the oddly shaped unit cell represents the trace on the inert substrate of the vertical symmetry plane between two adjacent sites (see Fig.SM1 in Supplemental Material for construction principle illustration). Hence, no-flux boundary condition can also be applied at the boundaries of the Voronoi unit cells analogously to the case of regular arrays [12].

In this situation Eq.(1) rewrites as:

$$i_{rand.arr.}(t) = \sum_n i_{cell}^{(n)}(t) \qquad (2)$$

where $i_{cell}^{(n)}(t)$ is the current in the *n*th unit cell and summation is taken over all of the Voronoi cells. Each $i_{cell}^{(n)}(t)$ value is formally unique since each cell is distinctive in terms of its shape, surface area and active sites position with respect to the boundaries and geometrical center of the cell. It is clear that even if such representation of the random array make use of local symmetries within the array it does not completely simplifies the simulations of the system since either whole array (or a sequence of all the Voronoi cells due to their independence) should be simulated. To overcome this difficulty one may employ approximation conventionally used for simulations of the electrochemical behavior of the regular arrays.

***Circular unit cell approximation.*** Considering mass transport in a unit cell of the squared or hexagonal regular array requires solving a 3D mathematical model. Nowadays it is easier than before, but still it is a cumbersome, resource and time consuming process especially when a large number of independent cells need to be modelled simultaneously. Therefore each original unit cell (i.e. squared or hexagonal one) is converted into a circular unit cell as shown in Fig.2 as was performed for regular arrays [1, 7]. This greatly facilitate simulations since now due to the axial symmetry of the domain the mass transport model can then be formulated in 2D cylindrical coordinates. Such formal conversion of the unit cell into a circular one evidently introduces a bias into the predicted responses of the unit cells. However, we demonstrated earlier that unless the unit cells are extremely small with respect to the active site size, the bias induced in the unit cell current due to transformation of its shape is less than few percent [13]. This validated the widely used cylindrical approximation. It should emphasized that the surface areas of the circular unit cells cross-sections must be equal to the area of the primary unit cell (see below "Chronoamperometric response of a unit cell"), which necessarily involves simultaneously non-negligible overlap and non-negligible avoided areas (Fig.2). The resulting bias can be evaluated for regular arrays (Fig.2a). This was shown to be non detrimental due to automatic compensation between overlapping and avoided solution volumes (Fig.2a, bottom).

This encouraged us [12] and Compton group [10, 11] to extend the circular unit cell approximation to the realm of randomly located active sites as exemplified in Fig.2b. It should be noted that the unit cells appear intersecting in the figure, nevertheless each of them acts independently due to the no-flux conditions applied at the side circular boundary of the cell as discussed above. Along with conversion of the unit cell shape the active sites were assumed to be at the center of each circular cell. This significantly reduced simulations computational cost while introducing only reasonable bias (1-7%) depending on the relative sizes of the cell and active site [12]. The latter is a good figure since experimental data not often available with accuracy better than 5% for such arrays. Note that several other approaches have been proposed for treating quantitatively electrochemical data gathered at random arrays although they are not relevant to the purpose of our former contributions [1, 7, 12-14] and those of Compton's group [10, 11]. See for example references [15-17].

In order to make the circular unit cell approximation fully operative, let us introduce dimensionless area of the Voronoi unit cell $\rho = A_{cell}/A_{site} = (r_{cell}/r_0)^2$, $A_{cell}$ and $A_{site}$ being geometric areas of the unit cell (being equal for original Voronoi cell and its circular approximation) and active site, $r_{cell}$ and $r_0$ are the unit cell and site radii. Then Eq.(2) rewrites as follows:

$$i_{rand.arr.}(t) = \sum_\rho i_{cell}^{(\rho)}(t) N_{site}^{(\rho)} \quad (3)$$

where summation is taken over all possible unit cells surface areas, $i_{cell}^{(\rho)}$ being the current in each circular unit cell of dimensionless area $\rho$ and $N_{site}^{(\rho)}$ the number of such cells. For a large array $\rho$ may be assumed to vary continuously (and not having just discrete values as in Eq.(3)). Thus, Eq.(3) can be rewritten in integral form as follows

$$i_{rand.arr.}(t) = N_{site} \int_1^\infty i_{cell}^{(\rho)}(t) f(\rho) \, d\rho \quad (4)$$

where $f(\rho)$ is a probability density of unit cell dimensionless area $\rho$ (in other words, $f(\rho)$ is a normalization of $N_{site}^{(\rho)}$ values).

Equation (4) is general and can be applied to very different random and regular arrays of large dimension. Figure 3 exemplifies two different cases: uniform random array (i.e. the active sites are equiprobably dispersed over the whole surface of the array according to uniform probability distribution) and Gaussian or binormal random array (i.e. sites have tendency to be located near geometric center of the array [left bottom corner in Fig.3b] and distributed according to 2D Gaussian distribution). Figs.3a,b display partial areas of each system for clarity of presentation, while the whole images of the respective arrays can be found in Supplemental Materials. As can be seen, $f(\rho)$ distributions for these two systems have similar structure, but the amplitude and shape are quite different ensuring different electrochemical responses of each of the array. It should be noted that in the case of regular arrays $f(\rho)$ distribution is delta-function, viz., $f(\rho) = \delta(\rho - \rho_0)$ since all the unit cells have the same dimensionless surface area $\rho_0$ and, hence, Eq.(4) simplifies into Eq.(1).

As soon as the $\rho$-distribution is known the response of an array can be evaluated via Eq.(4).

Additionally, since the unit cells have the same uniform circular structure except for their $\rho$ values the contributions $i_{cell}^{(\rho)}(t)$ can be evaluated and stored in advance or represented via analytical approximation further simplifying computations. To emphasize this important point, let us examine the chronoamperometric behavior of one of such unit cells.

***Chronoamperometric response of a unit cell.*** Considering diffusion limited conditions, as was indicated above, all unit cells experience two limiting Cottrell regimes irrespective of their sizes, although the time ranges in which this occurs depends on the unit cell dimensions. First, at short times (i.e. when $t \ll r_0^2/D$, $D$ is the diffusion coefficient of electroactive species) any unit cell gives rise to a Cottrell current governed by surface area $\pi r_0^2$, i.e.:

$$i_{cell}^0(t) = nFc_{bulk}r_0^2\sqrt{\pi D/t} \tag{5}$$

where $n$ is the number of the transferred electrons per elementary electron transfer (ET) step, $F$ is the Faraday constant, and $c_{bulk}$ the bulk concentration of electroactive species.

At larger times after beginning of the experiment ($t \gg r_{cell}^2/D$) the diffusion layer at each unit cell has fully developed (i.e. when the diffusion layers of adjacent unit cells fully overlap) thus creating a planar diffusion layer expanding into the solution [7]. Hence, a second Cottrell regime is observed with current proportional to the surface area of the unit cell:

$$i_{cell}^\infty(t) = nFc_{bulk}r_{cell}^2\sqrt{\pi D/t} \tag{6}$$

Since the same surface area of circular unit cell and the original Voronoi cell that it models must yield the same current contribution $i_{cell}^\infty(t)$, Eq.(6) provides the relationship between their surface areas.

The presence (or not) of a third regime depends on the relative size of the unit cell with respect to the size of an active site. Indeed, if the unit cell is sufficiently large its diffusion layer acquires a hemispherical shape before reaching the limit in Eq.(6) which results in a quasi-steady state current [7]. Thus for large unit cells ($\rho$ larger than few units) a perfectly defined transition takes place between the two above Cottrell regimes involving a quasi-steady state current, $i_{cell}^{ss} = 4nFDc_{bulk}r_0$. For smaller unit cells ($\rho \approx 1-3$) this regime cannot be fully observed resulting in a smooth transition between the two limiting currents.

These limiting regimes could be hardly identified at a conventional amperogram (Fig.4a). However, if the same amperogram is plotted in a log-log scale it reveals the two or three modes (Fig.4b) depending on $\rho$ value as just discussed. This representation of the current, which appears as a tilted sigmoid, suggests useful normalization for the problem at hand. Indeed, the current of a unit cell normalized with respect to its (time dependent) Cottrell limits:

$$\varphi_{cell}^\rho(t) = \frac{i_{cell}^\rho(t) - i_{cell}^0(t)}{i_{cell}^\infty(t) - i_{cell}^0(t)} \tag{7}$$

has a shape of a sigmoid (Fig.4c). It should be emphasized that hereafter this expression is used in the context of a circular unit cell for the sake of simplifying the presentation, however, the above analysis and normalization remain valid for original Voronoi cells with the same outcome [12, 14].

***Response of a random array.*** Normalization (7) has two immediate advantages. First, it can be extended thanks to Eq.(4) to provide the current of the whole random array in the normalized form (see Supplemental Material for the derivation of this expression) [12]:

$$\Phi(\tau) = \frac{i_{array}(\tau) - i^0_{array}(\tau)}{i^\infty_{array}(\tau) - i^0_{array}(\tau)} = \frac{1}{\rho_{avg}-1} \int_1^\infty \varphi^\rho_{cell}(\tau)(\rho-1)f(\rho)d\rho \qquad (8)$$

where $\tau = Dt/r_0^2$ is the dimensionless time, $i^0_{array}(\tau)$ and $i^\infty_{array}(\tau)$ are the corresponding Cottrell limits of the array. Second, due to the convenient shape of the dimensionless currents (7) they can be easily parametrized via rather simple analytical fitting equation [12]:

$$\varphi^\rho_{cell}(\tau) = \frac{1}{2}[1 + \tanh(b_0^\rho + b_1^\rho \ln(a\tau))] \qquad (9)$$

where $a = 9 \times 10^{-7}$ and parameters $b_0^\rho$ and $b_1^\rho$ are functions of unit cell surface area:

$$b_0^\rho = -2.866 + 35.383/\ln(11.543\,\rho)$$

$$b_1^\rho = 0.204 + 0.591/\ln(2.534\,\rho) \qquad (10)$$

as obtained through fitting of computed $\varphi^\rho_{cell}(\tau)$ responses for a large set of $\rho$ values. The approximation (9)-(10) is valid for $\rho \leq 10^3$, which is amply sufficient for our purpose (compare Figs.3c,d). Evidently, this can be extended to encompass a wider range of $\rho$ if necessary via the same principle.

Knowing the contributions $\varphi^\rho_{cell}(\tau)$ in analytical form allows a straightforward evaluation of random array responses from Eq.(8) as soon as their distributions of unit cell sizes $f(\rho)$ is available. The average value $\rho_{avg}$ does not need to be supplied as an independent entry since it can be readily computed from the $f(\rho)$ distribution according to the definition of a mean value:

$$\rho_{avg} = \int_1^\infty \rho f(\rho) d\rho \qquad (11)$$

Note that the lower integration limit in Eqs.(4) and (11) is unity since by construction $\rho \geq 1$.

Figure 5 displays the variations of the normalized currents evaluated via Eqs.(8)-(10) with the dimensionless time for the two types of random arrays shown in Fig.3. The average normalized surface areas for uniform and Gaussian arrays were $\rho^{uf}_{avg} = 23.2$ and $\rho^g_{avg} = 32.0$, correspondingly. In addition, the response of a regular array with $\rho^{reg}_{avg} = 25.0$ is provided for comparison.

Eq.(8) proves to be useful also for a more practical and complicated case of reconstruction the distribution of unit cell sizes on the basis of the measured current. Indeed analyzing $f(\rho)$ experimentally is a tedious and difficult exercise, while for some systems such characterization is even impossible. However, theoretically this can be achieved as follows based on the analysis of chronoamperometric currents. First, let us discretize $f(\rho)$ and represent it as a histogram, $H_M(\rho)$, with $M$ beans:

$$H_M(\rho) = \sum_{k=1}^{M} h_k B_k(\rho), \quad B_k(\rho) = \begin{cases} 1, & \rho_k - \Delta\rho/2 \leq \rho \leq \rho_k + \Delta\rho/2 \\ 0, & \text{otherwise} \end{cases} \quad (12)$$

where $B_k(\rho)$ is a bin function, $\rho_k$, $\Delta\rho$ and $h_k$ are center, width and height of a $k$-th bin. Vector of coefficients $h = (h_1, \ldots, h_M)$ represents thereby a distribution of the unit cell sizes of a random array as soon as the binning parameters ($\rho_k$, $\Delta\rho$) are set. Minimizing the difference between the evaluated and experimental currents upon varying the coefficients $h_k$ allows reconstruction of the thought distribution. The latter statement can be written formally as a constrained minimization problem:

$$\min_{h} [\Phi_{sim}(\tau, h) - \Phi_{exp}(\tau)]^2 \quad (13)$$

$$\sum_{k=1}^{M} h_k \Delta\rho = 1 \quad (14)$$

$$h_k \geq 0 \quad (15)$$

where $\Phi_{exp}$ is the normalized experimental current. The constraints (14)-(15) on coefficients $h_k$ follow from the properties of probability densities, more precisely that it is a non-negative function whose integral over its definition range is equal to unity.

For the sake of demonstration of described reconstructed procedure one may assume that the currents shown in Fig.5 are the experimental ones and the distribution in Fig.3c,d are not known. The corresponding $\rho$-distributions can then be attempted to be identified by solving optimization problem (13)-(15). The outcome of this is shown in Fig.6 revealing a very good agreement for both cases between the primary (solid lines) and reconstructed (histograms) distributions.

**Discussion**

As can be seen from Eq.(8) the distribution of unit cell sizes is the only characteristics responsible for unique features of electrochemical response of a given array as soon as the active sites activities and size are monodisperse. The variations of the normalized current in Fig.5 evaluated for different kind of arrays confirm this point. Indeed, the range displayed by the $\Phi(\tau)$ sigmoid along the time axis is largely defined by the average $\rho$ value. The larger average unit cell size the later is transition towards the second Cottrell regime, as evidenced by the shift to larger times of normalized current for Gaussian array (blue curve, Fig.5) relatively to uniform one (red curve, Fig.5) in agreement with their relative mean $\rho$ values $\rho_{avg}^{uf} = 23.2$ and $\rho_{avg}^{G} = 32.0$, respectively. Conversely, the dispersion of $\rho$-distribution affects the shape of the $\Phi(\tau)$. Regular arrays give rise to the fastest transition due to their high symmetry (black curve in Fig.5). Indeed, all their unit cells have the same sizes so that the transition between the different diffusion regimes occurs simultaneously for all of them. On the contrary, for the Gaussian array featuring a long tail in its $\rho$-distribution the transition is sluggish since unit cells corresponding to a range of larger $\rho$ perform their transitions to the second Cottrell regime at increasingly larger times. It should be noted that transition of the uniform random array starts earlier even if it is slower than the one of the regular array. This, in fact, reflects the unit cells with the smaller surface areas than those of the regular array (see Fig.6a), while the larger unit cells are responsible for sluggish behavior for $\Phi(\tau) > 0.4$ (Fig.5).

This sensitivity of $\Phi(\tau)$ to the shape of $\rho$-distribution make possible the reconstructions shown in Fig.6. Moreover, these results together with those obtained in our previous work [14] demonstrate that with this approach one may characterize, in a statistical sense, the distribution of the active sites and judge if the array under the scrutiny is regular or random as well as deduce information on the type of probability distribution governing the scattering of active sites over the inert substrate. This allows identifying possible biases introduced into the active sites distribution by fabrication procedure or other factors. Additional advantage of this framework that all this information can be extracted from chronoamperograms and does not require any sophisticated equipment, vacuum chambers etc.

It should be noted, however, that the reconstruction procedure described above relies on the normalization of the measured current (8), which suppose a good experimental resolution of each Cottrelian limiting regimes (5)-(6). This would be a rare case since the latter implies a good time resolution over several orders of time magnitude as can be seen from the time range in Fig.5. Nevertheless, as we demonstrated the reconstruction performs adequately well when realistic range of times are taken into account with proper approximations performed to renormalize the experimental current [14].

For the identification of the probability distribution governing locations of the active sites one may rely on evaluation of normalized currents with various $f(\rho)$ and then compare with the results of the reconstruction. In this respect, a good reference for this kind of comparison, consists in the empirical analytical distribution established and validated for random arrays with active sites distributed according to a uniform distribution as shown in Fig.3a,c (i.e. without any specific bias) [18]. For our case it writes as:

$$f_{gen}(\rho) = \frac{1}{\rho_{avg}} \frac{(7/2)^{7/2}}{\Gamma(7/2)} \times \left(\frac{\rho}{\rho_{avg}}\right)^{5/2} \exp\left[-\frac{7}{2}\frac{\rho}{\rho_{avg}}\right] \qquad (16)$$

where $\Gamma(x)$ is the Gamma function. It should be noted, however, that Eq.(16) is valid only in the limit of infinite number of sizeless active sites. In practice this corresponds to sufficiently large numbers of sites whose dimensions are significantly small with respect to the size of the system. This was the case for the array shown in Fig.3a (see Supplementing Material for more details). The solid line shown in Fig.3c was evaluated according to Eq.(16) and resulted in excellent agreement with the histogram of unit cell sizes obtained from this particular uniform random array.

In terms of perspective, our approach can be employed, in principle, for dynamic characterization of nanostructured electrodes with randomly distributed catalytic sites. For example, it is known that during their operation such systems often show some poisoning or clustering of the catalytic sites etc. Theoretical approach reviewed in this paper performed on a series of amperograms acquired sequentially during the system operation may evidence dynamical changes of catalytic sites with respect to their statistical distribution, average cluster size etc.

Finally, it should be stressed that the above described approach is valid under the assumptions indicated at the beginning of 'Results' section, that is all active sites have almost identical dimensions and perform in diffusion limiting regime (i.e. when the concentrations is constant across electrode-solution interface). In case of presence of significant ohmic drop, preceding or following to ET step chemical reactions (of the order higher than one) as well as slow charge transfer the concentrations at the electrode surface will be a function of position as well as the locations of the neighboring electrodes, thus resulting in different time- and space-competition between adjacent Voronoi cells. In simple terms, the pattern and sizes of the Voronoi cells defined at the electrode surface level will not be identical to those prevailing at different distances in the solution. Therefore, the pattern of Voronoi cells used in our works reviewed here and in those of Compton group will not be anymore a geometric characteristics of a given array, but will depend on the experimental time or scan rate for voltammetry as discussed in [12]. We are currently developing theoretical strategies, which seem to offer interesting solution to the problem. These will be reported in a future work after their validation, in particular in the context of voltammetry of fast and slow electron transfer couples.

**Conclusions**

The approach reviewed in this work allows an efficient modeling and prediction of the electrochemical behavior of nanostructured electrodes of identical sizes dispersed onto an inert substrate. It represents a unified framework to simulate electrochemical responses of regular and random arrays under diffusion limited conditions as soon as the distribution of unit cells dimensionless surface areas $f(\rho)$ is provided. Conversely, the latter can be reconstructed from chronoamperograms giving access to an extremely important information that is excessively difficult to obtain otherwise. In particular, the degree of active sites distribution randomness can be easily evaluated so that different types of random arrays can be differentiated.

**Acknowledgments**

This work was supported by CNRS, Ecole Normale Superieure – PSL Research University, Sorbonne University (UMR 8640 PASTEUR). CA acknowledges Xiamen University for his position of Distinguished Professor.

**Figure Captions**

**Figure 1.** Unit cells for active sites located on a square lattice (a) and top view of the Voronoi unit cells (b).

**Figure 2.** Schematic illustration of circular unit cell approximation for (a) regular array and (b) for random array. In each panel the bottom figure represents the cylindrical unit cells equivalents (a) or their cross-section with the inert substrate plane (b).

**Figure 3.** Statistically generated uniform (a) and Gaussian (b) random arrays (see text for the definitions) along with their Voronoi tesselations and their respective unit cell size distributions $f(\rho)$ (c, d). The average dimensionless surface areas are $\rho_{avg}^{uf} = 23.2$ (uniform) and $\rho_{avg}^{g} = 32.0$ (Gaussian), correspondingly. Solid lines in c) and d) are the fits of the histograms. Red line in c) evaluated via Eq.(16) (see below).

**Figure 4.** Amperometric responses of unit cells with $\rho = A_{cell}/A_{site} = 9, 25, 49$ and $100$ plotted in a) linear scale; b) log-log scale; c) normalized current via Eq.(7).

**Figure 5.** Normalized current $\Phi(t)$ for uniform (red, $\rho_{avg}^{uf} = 23.2$) and Gaussian (blue, $\rho_{avg}^{G} = 32.0$) random arrays shown in Fig.3 along with response of regular array (black, $\rho_{avg}^{reg} = 25.0$).

**Figure 6.** $f(\rho)$ distributions reconstructed (histograms) from the amperometric responses of (a) uniform and (b) Gaussian random arrays shown in Fig.5. The corresponding arrays are shown in Fig.3a,b and their original distributions (solid lines) are the same as shown in Fig. 3c,d.

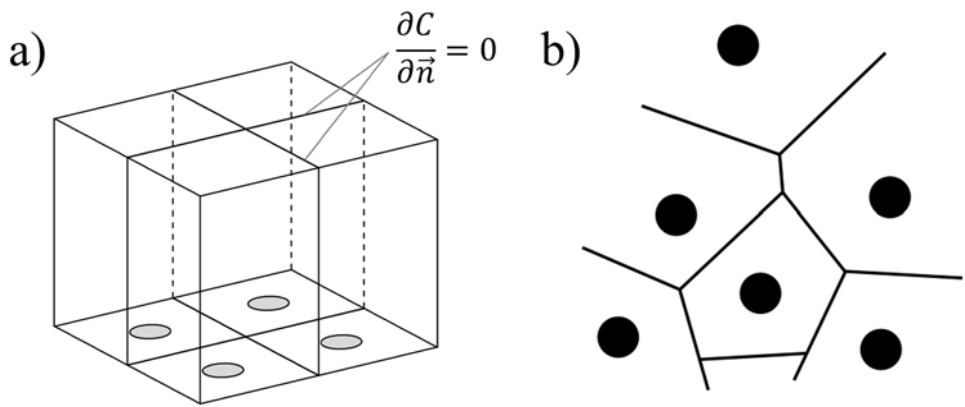

**Figure 1.** Unit cells for active sites located on a square lattice (a) and top view of the Voronoi unit cells (b).

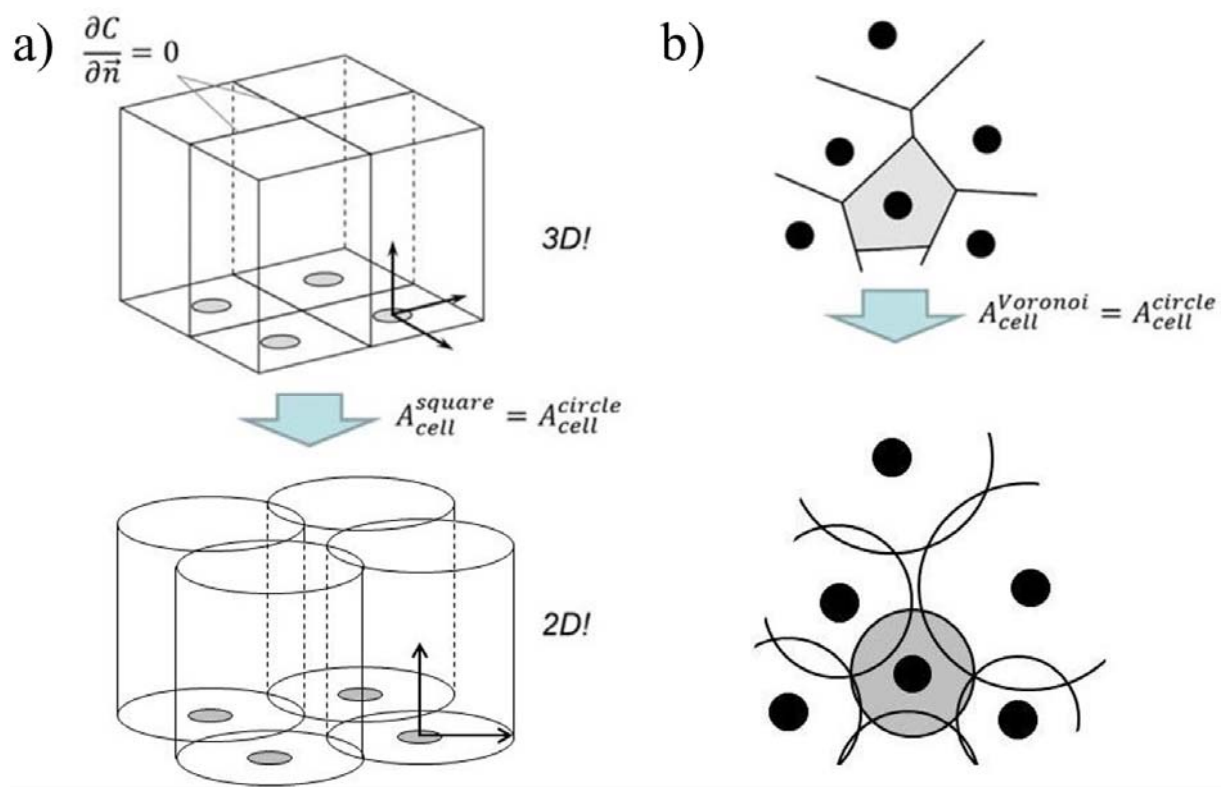

**Figure 2.** Schematic illustration of circular unit cell approximation for (a) regular array and (b) for random array. In each panel the bottom figure represents the cylindrical unit cells equivalents (a) or their cross-section with the inert substrate plane (b).

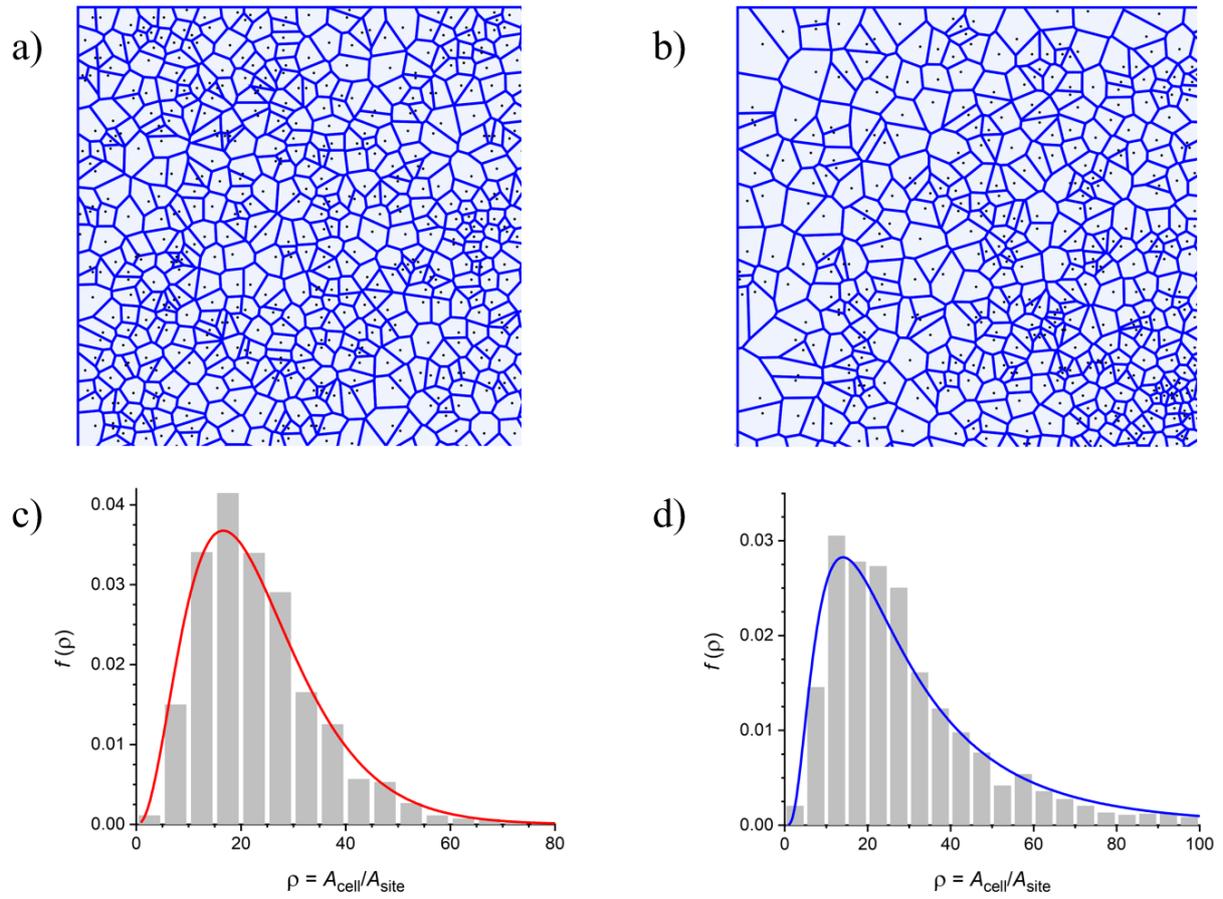

**Figure 3.** Statistically generated uniform (a) and Gaussian (b) random arrays (see text for the definitions) along with their Voronoi tesselations and their respective unit cell size distributions $f(\rho)$ (c, d). The average dimensionless surface areas are $\rho_{avg}^{uf} = 23.2$ (uniform) and $\rho_{avg}^{g} = 32.0$ (Gaussian), correspondingly. Solid lines in c) and d) are the fits of the histograms. Red line in c) evaluated via Eq.(16) (see below).

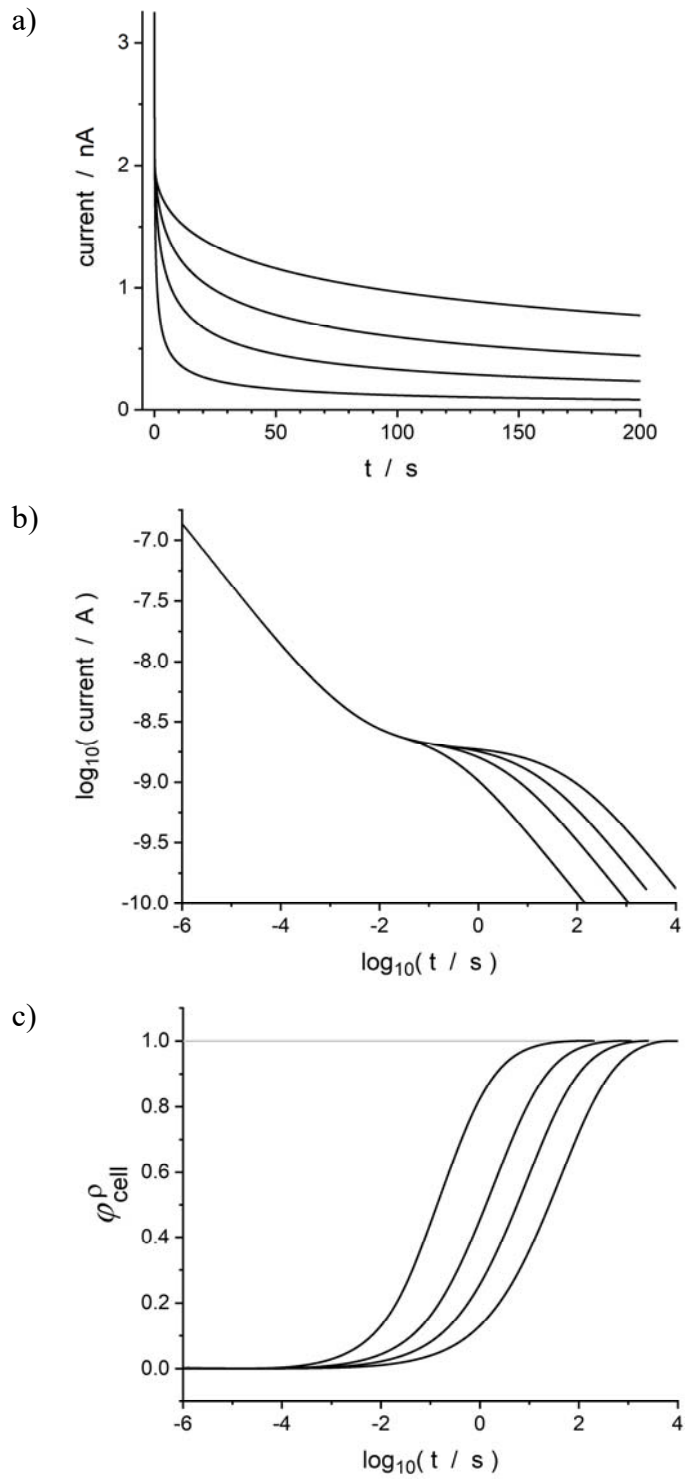

**Figure 4.** Amperometric responses of unit cells with $\rho = A_{cell}/A_{site} = 9, 25, 49$ and $100$ plotted in a) linear scale; b) log-log scale; c) normalized current via Eq.(7).

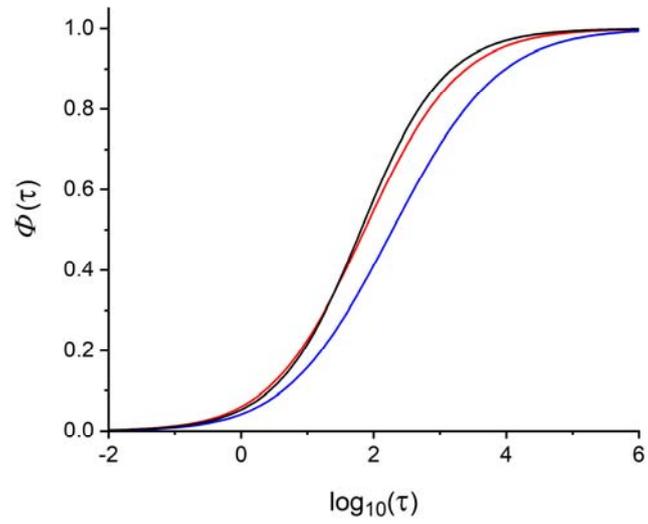

**Figure 5.** Normalized current $\Phi(t)$ for uniform (red, $\rho_{avg}^{uf} = 23.2$) and Gaussian (blue, $\rho_{avg}^{G} = 32.0$) random arrays shown in Fig.3 along with response of regular array (black, $\rho_{avg}^{reg} = 25.0$).

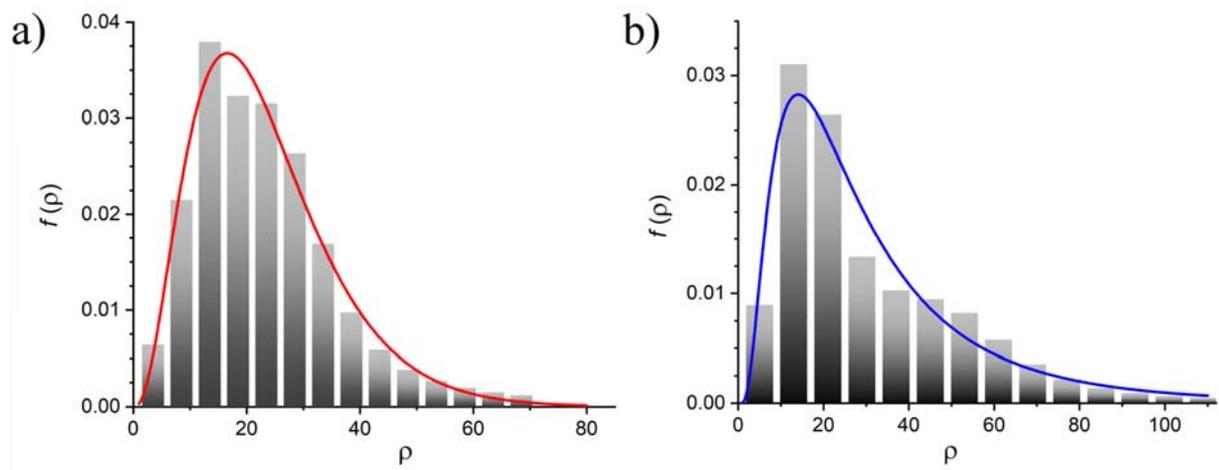

**Figure 6.** $f(\rho)$ distributions reconstructed (histograms) from the amperometric responses of (a) uniform and (b) Gaussian random arrays shown in Fig.5. The corresponding arrays are shown in Fig.3a,b and their original distributions (solid lines) are the same as shown in Fig. 3c,d.